\documentclass[prd,nofootinbib,showpacs,preprint,aps]{revtex4-1}

\usepackage{amsfonts}
\usepackage{mathrsfs}
\usepackage{epsfig}
\usepackage{graphicx}%
\usepackage{dcolumn}
\usepackage{amsmath}
\usepackage{color}
\usepackage{color}

\makeatletter
\def\btt#1{\texttt{\@backslashchar#1}}%
\DeclareRobustCommand\bblash{\btt{\@backslashchar}}%
\makeatother
\usepackage{subfigure}
\usepackage{amsmath}
\usepackage{amsfonts}
\usepackage{amssymb}

\usepackage[applemac]{inputenc}
\usepackage[T1]{fontenc}
\usepackage{amssymb}
\usepackage{amsfonts}
\usepackage{amsfonts}
\usepackage{graphicx}

\usepackage{bbm}

\begin{document}
\title{The metric of general rotating spacetimes}
\author{Changjun Gao}
\email{gaocj@nao.cas.cn}
\affiliation{National Astronomical Observatories, Chinese Academy of Sciences, 20A Datun Road, Beijing 100101, China}

\begin{abstract}\baselineskip=18pt
We propose the metric for general rotating spacetimes. These spacetimes are stationary, axially symmetric and asymptotically flat in space. They can be the spacetimes outside of rotating black holes or rotating celestial bodies such as the Sun and the Earth. The metric functions are expanded in power series of distance. The angle variable appears in the expansion coefficients with the form of powers of $\cos\theta$. 
\vskip 1in

\end{abstract}

\maketitle

\section{introduction}
The Einstein field equations are a set of nonlinear
partial differential equations. Therefore, it is difficult to find the exact solutions. In view of this point, people turned to other ways instead of struggling with the Einstein equations, e.g. the
Newman-Janis algorithm by using of the complex transformation \cite{new:2014}, the Newman-Penrose formalism by using of the spinors \cite{penrose:1984},
and the method of Backlund transformations \cite{har:1978}. Despite their great success in
dealing with Einstein equations, these methods are
technically complicated and expert-oriented.

In this study, we present an ansatz for the parameterizations of metric for general rotating spacetimes. Since it is the most general rotating spacetime, it is applicable to any theories of modified gravity. The ansatz originates from the idea of Taylor expansion and similar in spirit to
the parameterized post-Newtonian approach (PPN) which describes
the spacetime far away from the source \cite{will:2006}. One of the first parametrization for black holes was
proposed by Johannsen and Psaltis \cite{job:2011}. They expressed the metric functions in terms of a Taylor expansion in
powers of Newtonian potential. However, as
pointed out by Refs. \cite{car:2014,kono:2016}, these parameterizations face a number of
difficulties: (i) There are infinite number
of roughly equally important parameters which make the parameterizations
difficult to isolate the dominant ones. (ii) The parametrization can only be employed in the small
deviations from General Relativity, but fails for other modified gravities. (iii) The adoption of the Janis-Newman transformations
does not allow one to reproduce alternatives to the Kerr
spacetime in alternatives of gravity theory.

For that reason, the Johannsen-Psaltis approach does
not seem to be a robust and generic parametrization for rotating
black holes.
Considering this, Roman Konoplya, Luciano Rezzolla and Alexander Zhidenko proposed a parametric framework to
describe the spacetime of axisymmetric black holes in generic metric theories of gravity. In particular, they used a continued-fraction expansion in terms
of the radial distance. At the same time, a Taylor expansion in
terms of the cosine function of the angular variable is taken.  On the other hand, Delaporte, Eichhorn and Held \cite{dela:2022} discussed the
parameterizations of black-hole spacetimes in and beyond General Relativity and showed that the existing parameterizations are included in their new parameterizations.

In this work, we report a novel ansatz for the metric of general rotating spacetimes. The metric is motivated by the expansion of Kerr metric in isotropic coordinates. In detail, the metric functions are expanded into the sum of gravitational multipolar moments. This is different from the Refs. \cite{job:2011,kono:2016,dela:2022}.

\section{Kerr spacetime in isotropic coordinates}
The Kerr spacetime in isotropic coordinates reads
\begin{eqnarray}
ds^2=g_{00}dt^2+g_{11}\left(dr^2+r^2d\theta^2\right)+g_{33}r^2\sin^2\theta d\phi^2+2g_{03}r\sin^2\theta dtd\phi\;,
\end{eqnarray}
with
\begin{eqnarray}
&&g_{00}=-\frac{a^4-8a^2r^2-2a^2M^2+\left(4r^2-M^2\right)^2+16a^2r^2\cos^2\theta}{
a^4-8a^2r^2-8Mra^2-2a^2M^2+\left(2r+M\right)^4+16a^2r^2\cos^2\theta^2}\;,\\
&&g_{11}=\frac{a^4-8a^2r^2-8a^2rM-2a^2M^2+\left(2r+M\right)^4+16a^2r^2\cos^2\theta}{16r^4}\;,\\
&&g_{03}=-\frac{8\left(-a^2+4r^2+4rM+M^2\right)aM}{a^4-8a^2r^2-8a^2rM-2a^2M^2+\left(2r+M\right)^4+16a^2r^2\cos^2\theta}\;,\\
&&g_{33}=\left[a^2\left(a^2+8r^2-8rM-2M^2\right)+\left(2r+M\right)^4+4ar\left(a^2+4r^2-M^2\right)\sin\theta
\right]\nonumber\\&&
\times \frac{\left[a^2\left(a^2+8r^2-8rM-2M^2\right)+\left(2r+M\right)^4-4ar\left(a^2+4r^2-M^2\right)\sin\theta
\right]}{
16r^4\left[a^2\left(a^2+8r^2-8rM-2M^2\right)+\left(2r+M\right)^4+16a^2r^2\cos^2\theta\right]}\;.
\end{eqnarray}
Here $M$ is the mass and $a$ the angular momentum per unit mass, respectively, for the black hole.
Expanding the metric functions into power series of inverse $r$, we obtain
\begin{eqnarray}
&&g_{00}=-1+\frac{2M}{r}-\frac{2M^2}{r^2}+\frac{M}{2}\cdot\frac{a^2+3M^2-4a^2\cos^2\theta}{r^3}+O\left(\frac{1}{r^4}\right)\;,\\
&&g_{11}=1+\frac{2M}{r}-\frac{1}{2}\cdot\frac{3M^2-a^2+2a^2\cos^2\theta}{r^2}+\frac{1}{2}\cdot\frac{M^3-Ma^2}{r^3}+\frac{\left(M^2-a^2\right)^2}{16r^4}\;,\\
&&g_{33}=1+\frac{2M}{r}-\frac{1}{2}\cdot\frac{3M^2+a^2}{r^2}+\frac{M}{2}\cdot\frac{3a^2+M^2-4a^2\cos^2\theta}{r^3}+O\left(\frac{1}{r^4}\right)\;,\\
&&g_{03}=-\frac{2Ma}{r^2}+\frac{2M^2a}{r^3}+\frac{\frac{1}{2}Ma\left(-a^2-3M^2+4a^2\cos^2\theta\right)}{r^4}+O\left(\frac{1}{r^5}\right)\;.
\end{eqnarray}
Observing these expansions, we find there are two parameters $M$ and $a$ in the expressions. In the natural units, $M$, $a$ and $r$ have the same dimension. What is more, the function of $\cos\theta$ (and its powers) appears in the numerators. Theses metric functions are the power series of inverse $r$ and the angular variable $\theta$ is included in the expansion coefficients. These expansions are instructive since they clearly show the presence of Newtonian gravitational potential, the gravitational quadrupole, octupole and other higher moments for the
mass distribution. However, it is subtle that all the multipolar moments are highly
coupled to each other in the Kerr solution. In other words, the multipolar moments are not independent free parameters but determined by only two free parameters, the mass $M$ and the specific angular momentum $a$. But in the case of real physical stars, the
gravitational quadrupole, octupole and higher moments for the
mass distribution are in principle independently specified. For example, the Earth has the shape of a pear. We need more free parameters to describe the spacetime outside of the Earth except for the mass $M$ and the specific angular momentum $a$. Therefore, the Kerr solution cannot describe the exterior spacetime of the Earth.  Of course, in practice the higher pole fields fall off very quickly such that the lowest multipolar dominates far away from the Earth. It is in the sense of asymptotics that the Kerr solution describes the exterior spacetime of a rotating object. In the light of this, we shall seek for the general rotating spacetimes among which the Kerr solution turns out to be a particular one.
\section{Parameterisations of general rotating spacetimes}
According to the usual Lewis-Papapetrou ansatz \cite{wald:1984} for a stationary and axially symmetric, rotating spacetime with two
Killing vector fields $\xi=\partial_{t}\;, \chi=\partial_{\phi}$, the corresponding metric in isotropic coordinates reads \cite{klei:2001}
\begin{eqnarray}
ds^2=-fdt^2+\frac{m}{f}\left(dr^2+r^2d\theta^2\right)+\frac{l}{f}r^2\sin^2\theta\left(d\phi-\frac{\omega}{r}dt\right)^2\;,
\end{eqnarray}
where $f,\ m,\ l,\ \omega$ are functions of only $r$ and $\theta$. This ansatz satisfies the circularity
and Frobenius conditions \cite{wald:1984,papa:1969,heu:1996}. The boundary conditions at infinity, $f=m=l=1,\ \omega=0$, ensure that the solutions are
asymptotically flat in space.

The global charges of the spacetime are determined from their asymptotic
behaviour. In particular, expansion at infinity yields \cite{isl:1985,klei:2001}
\begin{eqnarray}
f\longrightarrow 1-\frac{2M}{r}\;,\ \ \ \omega\longrightarrow \frac{2Ma}{r^2}\;,
\end{eqnarray}
determining the mass $M=\frac{1}{2}\lim_{r\rightarrow\infty}r^2\partial_r f$ and the angular momentum $J=\frac{1}{2}\lim_{r\rightarrow\infty}r^2\omega$.
Given the expansions of Kerr solution, we parameterize the metric functions as follows

\begin{eqnarray}\label{metric}
l&=&1+\frac{1}{r}\cdot \left(a_{1\bar{0}}+a_{1\bar{1}}\cos\theta+a_{1\bar{2}}\cos^2\theta+a_{1\bar{3}}\cos^3\theta+\cdot\cdot\cdot\right)\nonumber\\&&+\frac{1}{r^2}\cdot\left(a_{2\bar{0}}+a_{2\bar{1}}
\cos\theta+a_{2\bar{2}}\cos^2\theta+a_{2\bar{3}}\cos^3\theta+\cdot\cdot\cdot\right)\nonumber\\&&+\frac{1}{r^3}\cdot\left(a_{3\bar{0}}+a_{3\bar{1}}\cos\theta+a_{3\bar{2}}\cos^2\theta+a_{3\bar{3}}
\cos^3\theta+\cdot\cdot\cdot\right)+\cdot\cdot\cdot\;,\\
m&=&1+\frac{1}{r}\cdot \left(b_{1\bar{0}}+b_{1\bar{1}}\cos\theta+b_{1\bar{2}}\cos^2\theta+b_{1\bar{3}}\cos^3\theta+\cdot\cdot\cdot\right)\nonumber\\&&+\frac{1}{r^2}\cdot\left(b_{2\bar{0}}
+b_{2\bar{1}}
\cos\theta+b_{2\bar{2}}\cos^2\theta+b_{2\bar{3}}\cos^3\theta+\cdot\cdot\cdot\right)\nonumber\\&&+\frac{1}{r^3}\cdot\left(b_{3\bar{0}}+b_{3\bar{1}}\cos\theta+a_{3\bar{2}}
\cos^2\theta+b_{3\bar{3}}
\cos^3\theta+\cdot\cdot\cdot\right)+\cdot\cdot\cdot\;,\\
f&=&1+\frac{1}{r}\cdot \left(c_{1\bar{0}}+c_{1\bar{1}}\cos\theta+c_{1\bar{2}}\cos^2\theta+c_{1\bar{3}}\cos^3\theta+\cdot\cdot\cdot\right)\nonumber\\&&+\frac{1}{r^2}\cdot\left(c_{2\bar{0}}+c_{2\bar{1}}
\cos\theta+c_{2\bar{2}}\cos^2\theta+c_{2\bar{3}}\cos^3\theta+\cdot\cdot\cdot\right)\nonumber\\&&+\frac{1}{r^3}\cdot\left(c_{3\bar{0}}+c_{3\bar{1}}\cos\theta
+c_{3\bar{2}}\cos^2\theta+c_{3\bar{3}}
\cos^3\theta+\cdot\cdot\cdot\right)+\cdot\cdot\cdot\;,\\
\omega&=&\frac{1}{r^2}\left[d_{0}+\frac{1}{r}\cdot \left(d_{1\bar{0}}+d_{1\bar{1}}\cos\theta+d_{1\bar{2}}\cos^2\theta+d_{1\bar{3}}\cos^3\theta+\cdot\cdot\cdot\right)\right.\nonumber\\&&\left.+\frac{1}{r^2}\cdot
\left(d_{2\bar{0}}+c_{2\bar{1}}
\cos\theta+d_{2\bar{2}}\cos^2\theta+d_{2\bar{3}}\cos^3\theta+\cdot\cdot\cdot\right)\right.\nonumber\\&&\left.+\frac{1}{r^3}\cdot\left(d_{3\bar{0}}+d_{3\bar{1}}\cos\theta
+d_{3\bar{2}}\cos^2\theta+c_{3\bar{3}}
\cos^3\theta+\cdot\cdot\cdot\right)+\cdot\cdot\cdot\;,\right]\;.
\end{eqnarray}

These parameterizations satisfy the boundary conditions $\partial_\theta l=\partial_\theta m=\partial_\theta f=\partial_\theta \omega=0$ on the symmetry axis ($\theta=0$) which is imposed by axial symmetry and regularity of the spacetime. The absence of conical singularities implies
also $l=m$ at $\theta=0$. As mentioned earlier, these expansions clearly illustrate the terms of gravitational multipolar moments.

Then the vacuum Einstein equations $G_{\mu\nu}=0$ give the nonvanishing coefficients as presented in the Appendix A.
Then we obtain
\begin{eqnarray}
g_{00}&=&-f+\frac{l\omega^2\sin^2\theta}{f}=-1-\frac{c_{1\bar{0}}}{r}-\frac{\frac{1}{2}c_{1\bar{0}}^2+c_{2\bar{1}}\cos\theta}{r^2}
\nonumber\\&&+\frac{\left(\frac{1}{2}c_{1\bar{0}}a_{2\bar{0}}+3c_{3\bar{0}}-\frac{1}{2}c_{1\bar{0}}^3\right)\cos^2\theta-c_{1\bar{0}}c_{2\bar{1}}\cos\theta-c_{3\bar{0}}}{r^3}+\cdot\cdot\cdot\;,\\
g_{11}&=&\frac{m}{f}=1-\frac{c_{1\bar{0}}}{r}+\frac{\frac{1}{4}c_{1\bar{0}}^2-a_{2\bar{0}}-c_{2\bar{1}}\cos\theta+\left(\frac{1}{4}c_{1\bar{0}}^2+2a_{2\bar{0}}\right)\cos^2\theta}{r^2}
\nonumber\\&&+\frac{1}{r^3}\cdot\left[\left(4a_{3\bar{1}}+c_{1\bar{0}}c_{2\bar{1}}\right)\cos^3\theta-\left(\frac{3}{2}c_{1\bar{0}} a_{2\bar{0}}-3c_{3\bar{0}}+\frac{3}{4}c_{1\bar{0}}^3\right)\cos^2\theta-3a_{3\bar{1}}\cos\theta\right.\nonumber\\&&\left.
-c_{3\bar{0}}+\frac{1}{4}c_{1\bar{0}}^3+c_{1\bar{0}}a_{2\bar{0}}\right]+\cdot\cdot\cdot\;,\\
g_{33}&=&\frac{l}{f}=1-\frac{c_{1\bar{0}}}{r}+\frac{\frac{1}{2}c_{1\bar{0}}^2+a_{2\bar{0}}-c_{2\bar{1}}\cos\theta}{r^2}+\frac{1}{r^3}\cdot\left[\left(\frac{1}{2}c_{1\bar{0}
}a_{2\bar{0}}+3c_{3\bar{0}}
-\frac{1}{2}c_{1\bar{0}}^3\right)\cos^2\theta\right.\nonumber\\&&\left.+\left(c_{1\bar{0}} c_{2\bar{1}}+a_{3\bar{1}}\right)\cos\theta-c_{3\bar{0}}-a_{2\bar{0}}c_{1\bar{0}}\right]
+\cdot\cdot\cdot\;,\\
g_{03}&=&-\frac{l\omega}{f}=-\frac{d_0}{r^2}-\frac{\frac{1}{2}{d_0c_{1\bar{0}}}+d_{1\bar{1}}\cos\theta}{{r^3}}\nonumber\\&&
+\frac{\left(5d_{2\bar{0}}-6d_0c_{1\bar{0}}^2+\frac{9}{2}d_0a_{2\bar{0}}\right)\cos^2\theta-\left(d_0c_{2\bar{1}}+\frac{1}{3}c_{1\bar{0}}d_{1\bar{1}}\right)
\cos\theta-d_{2\bar{0}}-d_0a_{2\bar{0}}+d_0c_{1\bar{0}}^2}{r^4}\nonumber\\&&+\cdot\cdot\cdot\;.
\end{eqnarray}
They describe the spacetime outside any finite, axially symmetric and rotating objects. By choosing different set of constant coefficients, we would obtain different rotating spacetime for vacuum Einstein equations. For example, if we fix
\begin{eqnarray}
&&c_{1\bar{0}}=-2M\;,\ \ \ \ d_0=2Ma\;,
\end{eqnarray}
and express the other coefficients with $M$ and $a$ as follows,
\begin{eqnarray}\label{kerr}
&&a_{2\bar{0}}=\frac{1}{2}\left(a^2-M^2\right)\;,\ \ \ \ c_{2\bar{1}}=0\;,\ \ \ \ c_{3\bar{0}}=-\frac{1}{2}M\left(a^2+3M^2\right)\;,\ \ \ \ \  \cdot\cdot\cdot\cdot\cdot\cdot\;,
\end{eqnarray}
we would obtain the Kerr spacetime. As noted, all the multipolar moments are subtly and highly
related to each other in the Kerr solution. Therefore, the Kerr spacetime is not for a real rotating celestial body, but for a rotating black hole.

At such a time, we had better make some supplementary explanations on Kerr solution. Birkhoff theorem tells us that any spherically symmetric solution to the vacuum Einstein equations must be static and asymptotically flat. This means that the exterior solution (i.e. the spacetime outside of a spherical, non-rotating, gravitating body) must be given by the Schwarzschild metric. However, the situation is complicated for rotating spacetimes. In the first place, we should emphasize that there is actually no Birkhoff-like theorem for axisymmetric spacetimes in General Relativity. In fact, it is not the truth that the spacetime in the vacuum region outside a generic rotating star (or the earth and
planet) must be a part of the Kerr spacetime \cite{david:2009,kerr:1965}. The much
milder statement is that outside a rotating star (or a planet)
the spacetime asymptotically approaches the Kerr solution \cite{jozef:2009}.
A remarkable fact for this point is that in the Kerr solution all the multipolar moments are highly
related to each other. In contrast, the solutions (16-19) allow many free parameters such as the mass $c_{1\bar{0}}$, the angular momentum $d_0$, the ``gravitational dipole'' $c_{2\bar{1}}$, the gravitational quadrupole $c_{2\bar{0}}$ and so on. In this sense, the metric given in Eqs.~(16-19) describe the spacetime of vacuum region outside any finite, axially symmetric and rotating objects.

\section{rotating spacetime with scalar hair}
The parameterizations for rotating spacetimes can be applied to non-vacuum Einstein equations and the modified theories of gravity. In this section, we consider the Einstein equations with the energy-momentum tensor of massless scalar field as an example. The Einstein equations with massless scalar field and the equation of motion for the scalar can be written as
\begin{eqnarray}
R_{\mu\nu}=\frac{1}{2}\nabla_{\mu}\varphi\nabla_{\nu}\varphi\;,\ \ \ \ \ \ \ \ \nabla_{\mu}\nabla^{\mu}\varphi=0\;,
\end{eqnarray}
with $\varphi$ is the scalar field. In order to obtain an asymptotically flat spacetime, we can parameterize the scalar field as follows
\begin{eqnarray}
\varphi&=&\frac{q_{1\bar{0}}+q_{1\bar{1}}\cos\theta+q_{1\bar{2}}\cos^2\theta+q_{1\bar{3}}\cos^3\theta+\cdot\cdot\cdot}{r}
\nonumber\\&&+\frac{q_{2\bar{0}}+q_{2\bar{1}}\cos\theta+q_{2\bar{2}}\cos^2\theta+q_{2\bar{3}}\cos^3\theta+\cdot\cdot\cdot}{r^2}\nonumber\\&&+\frac{q_{3\bar{0}}
+q_{3\bar{1}}\cos\theta+q_{3\bar{2}}\cos^2\theta+q_{3\bar{3}}\cos^3\theta+\cdot\cdot\cdot}{r^3}+\cdot\cdot\cdot\;,
\end{eqnarray}
and the metric functions are parameterized as given by Eqs.~(12-15). We find from the Einstein equations and the equation of motion for the scalar that
\begin{eqnarray}
g_{00}&=&-f+\frac{l\omega^2\sin^2\theta}{f}=-1-\frac{c_{1\bar{0}}}{r}-\frac{\frac{1}{2}c_{1\bar{0}}^2+c_{2\bar{1}}\cos\theta}{r^2}
\nonumber\\&&+\frac{\left(\frac{1}{2}c_{1\bar{0}}a_{2\bar{0}}+3c_{3\bar{0}}-\frac{1}{2}c_{1\bar{0}}^3\right)\cos^2\theta-c_{1\bar{0}}c_{2\bar{1}}\cos\theta-c_{3\bar{0}}}{r^3}+\cdot\cdot\cdot\;,\\
g_{11}&=&\frac{m}{f}=1-\frac{c_{1\bar{0}}}{r}+\frac{\frac{1}{4}c_{1\bar{0}}^2-a_{2\bar{0}}-\frac{1}{16}q_{1\bar{0}}^2-c_{2\bar{1}}\cos\theta+\left(\frac{1}{4}
c_{1\bar{0}}^2+2a_{2\bar{0}}+\frac{1}{16}q_{1\bar{0}}^2\right)\cos^2\theta}{r^2}
\nonumber\\&&+\frac{1}{r^3}\cdot\left[\left(4a_{3\bar{1}}+c_{1\bar{0}}c_{2\bar{1}}\right)\cos^3\theta-\left(\frac{3}{2}c_{1\bar{0}} a_{2\bar{0}}-3c_{3\bar{0}}+\frac{3}{4}c_{1\bar{0}}^3+\frac{1}{16}c_{1\bar{0}}q_{1\bar{0}}^2\right)\cos^2\theta-3a_{3\bar{1}}\cos\theta\right.\nonumber\\&&\left.
-c_{3\bar{0}}+\frac{1}{4}c_{1\bar{0}}^3+c_{1\bar{0}}a_{2\bar{0}}+\frac{1}{16}c_{1\bar{0}}q_{1\bar{0}}^2\right]+\cdot\cdot\cdot\;,\\
g_{33}&=&\frac{l}{f}=1-\frac{c_{1\bar{0}}}{r}+\frac{\frac{1}{2}c_{1\bar{0}}^2+a_{2\bar{0}}-c_{2\bar{1}}\cos\theta}{r^2}+\frac{1}{r^3}\cdot\left[\left(\frac{1}{2}c_{1\bar{0}
}a_{2\bar{0}}+3c_{3\bar{0}}
-\frac{1}{2}c_{1\bar{0}}^3\right)\cos^2\theta\right.\nonumber\\&&\left.+\left(c_{1\bar{0}} c_{2\bar{1}}+a_{3\bar{1}}\right)\cos\theta-c_{3\bar{0}}-a_{2\bar{0}}c_{1\bar{0}}\right]
+\cdot\cdot\cdot\;,\\
g_{03}&=&-\frac{l\omega}{f}=-\frac{d_0}{r^2}-\frac{\frac{1}{2}{d_0c_{1\bar{0}}}+d_{1\bar{1}}\cos\theta}{{r^3}}\nonumber\\&&
+\frac{\left(5d_{2\bar{0}}-6d_0c_{1\bar{0}}^2+\frac{9}{2}d_0a_{2\bar{0}}\right)\cos^2\theta-\left(d_0c_{2\bar{1}}+\frac{1}{3}c_{1\bar{0}}d_{1\bar{1}}\right)
\cos\theta-d_{2\bar{0}}-d_0a_{2\bar{0}}+d_0c_{1\bar{0}}^2}{r^4}+\cdot\cdot\cdot\;,\\
\varphi&=&\frac{q_{1\bar{0}}}{r}-\frac{\frac{1}{6}a_{2\bar{0}}q_{1\bar{0}}}{r^3}+\cdot\cdot\cdot\;,
\end{eqnarray}
with the nonvanishing coefficients given in Appendix B. They constitute the most general stationary, axially symmetric and asymptotically flat solution to the Einstein equations with massless scalar field. For comparison, Janis-Newman-Winicour \cite{JNW:1968} obtained the most general static spherically symmetric and asymptotically flat solution to the Einstein-massless-scalar field equations which has the line element as follows \cite{JNW:1968,virb:1997}
\begin{eqnarray}
ds^2&=&-\left(1-\frac{2M\sqrt{1+\eta^2}}{r}\right)^{\frac{1}{\sqrt{1+\eta^2}}}dt^2+\left(1-\frac{2M\sqrt{1+\eta^2}}{r}\right)^{-\frac{1}{\sqrt{1+\eta^2}}}dr^2\nonumber\\&&
+\left(1-\frac{2M\sqrt{1+\eta^2}}{r}\right)^{1-\frac{1}{\sqrt{1+\eta^2}}}r^2\left(d\theta^2+\sin^2\theta d\phi^2\right)\;,
\end{eqnarray}
with $\eta$ defined as the ratio of ``scalar charge'' $q$ to the black hole mass $M$,
\begin{eqnarray}
\eta\equiv\frac{q}{M}\;.
\end{eqnarray}
The scalar field is given by
\begin{eqnarray}
\phi=\frac{\eta}{\sqrt{1+\eta^2}}\ln\left(1-\frac{2M\sqrt{1+\eta^2}}{r}\right)\;.
\end{eqnarray}
There are two parameters, the black hole mass $M$ and the ``scalar charge'' $q$ for the solution. By switching off the ``scalar charge'' one recovers the well-known Schwarzschild solution. Compared with the static solution, the rotating one has richer physical content. This is signified by the vast number of independent and free parameters, for instance, the mass $c_{1\bar{0}}$, the angular momentum $d_0$, the scalar hair $q_{1\bar{0}}$ and so on.

\section{conclusion and discussion}
We know that if the scale of the object we observe, is much smaller than the distance to us, the spacetime far away from the object is perfectly described by the Schwarzschild spacetime. That is because the effect of rotation decays faster than the mass with the increase of distance. But with the decrease of distance, we would find the Lense-Thirring metric \cite{lense:1918} which is for a slowly rotating point mass describes the object very well compared to the Schwarzschild metric. If we come still nearer, we would detect the effects of the scalar hair, the gravitational quadrupole, octupole and so on if these do exist in the celestial body. We believe the general rotating solution derived in the above sections is applicable to the observations of the actual celestial bodies from far to near.

As a comparison, the gravitational potential outside of the Earth's (or other celestial object's) can be described by series of spherical harmonics in Newtonian gravity. Of course, the gravitational potential outside of any shape of objects
can be represented by a series of spherical harmonics in Newtonian gravity. That is because the corresponding gravitational potentials satisfy the Laplace equation according to Newtonian gravity. However, it is not the case in General Relativity. Different from the Laplace equation, the Einstein equations are nonlinear. So the metric functions are in general not determined by linear Laplace equations such that they are not allowed to be expanded into series of spherical harmonics as shown in Eqs.~(12-15).

\section{Acknowledgments}
This work is partially supported by the NSFC under grants 11633004 and 11773031.

\appendix

\section{the nonvanishing coefficients for vacuum Einstein equations}
\label{AppendixA}
\begin{eqnarray}
&&d_{1\bar{0}}=\frac{3}{2}c_{1\bar{0}}d_0\;,\ \ \ b_{2\bar{0}}=-\frac{1}{4}c_{1\bar{0}}^2-a_{2\bar{0}}\;,\ \ \ b_{2\bar{2}}=\frac{1}{4}c_{1\bar{0}}^2+2a_{2\bar{0}}\;,\ \ c_{2\bar{0}}=\frac{1}{2}c_{1\bar{0}}^2\;,\ \ d_{2\bar{1}}=2d_0c_{2\bar{1}}+\frac{4}{3}c_{1\bar{0}}d_{1\bar{1}}\;,\nonumber\\&& d_{2\bar{2}}=-5d_{2\bar{0}}+6d_0c_{1\bar{0}}^2-\frac{9}{2}d_0a_{2\bar{0}}\;,\ \ \ b_{3\bar{1}}=-3a_{3\bar{1}}-c_{1\bar{0}}c_{2\bar{1}}\;,\ \
b_{3\bar{3}}=4a_{3\bar{1}}+c_{1\bar{0}}c_{2\bar{1}}\;,\ \ \  c_{3\bar{1}}=c_{1\bar{0}}c_{2\bar{1}}\;,\nonumber\\&& c_{3\bar{2}}=\frac{1}{2}c_{1\bar{0}}^3-\frac{1}{2}c_{1\bar{0}}a_{2\bar{0}}-3c_{3\bar{0}}\;,\ \ \ d_{3\bar{0}}=-\frac{1}{4}c_{1\bar{0}}d_0a_{2\bar{0}}+\frac{7}{4}d_0c_{3\bar{0}}+\frac{5}{4}c_{1\bar{0}}d_{2\bar{0}}-\frac{9}{8}c_{1\bar{0}}^3d_0-\frac{1}{12}
c_{2\bar{1}}d_{1\bar{1}}\;,\nonumber\\&& d_{3\bar{2}}=\frac{69}{8}c_{1\bar{0}}^3d_0-\frac{27}{4}c_{1\bar{0}}d_0a_{2\bar{0}}-\frac{25}{4}c_{1\bar{0}}d_{2\bar{0}}-\frac{27}{4}d_0c_{3\bar{0}}+\frac{7}{4}c_{2\bar{1}}
d_{1\bar{1}}\;,\nonumber\\&& d_{3\bar{3}}= \frac{22}{3}c_{1\bar{0}}d_0c_{2\bar{1}}-\frac{7}{3}d_{3\bar{1}}-\frac{9}{4}d_0a_{3\bar{1}}-2a_{2\bar{0}}d_{1\bar{1}}+\frac{20}{9}c_{1\bar{0}}^2d_{1\bar{1}}\;,\ \ \ \ \cdot\cdot\cdot\cdot\cdot\cdot\;.
\end{eqnarray}
\section{the nonvanishing coefficients for Einstein-massless-scalar equations}
\label{AppendixB}
\begin{eqnarray}
&&d_{1\bar{0}}=\frac{3}{2}c_{1\bar{0}}d_0\;,\ \ \ b_{2\bar{0}}=-\frac{1}{4}c_{1\bar{0}}^2-a_{2\bar{0}}-\frac{1}{16}q_{1\bar{0}}^2\;,\ \ \ b_{2\bar{2}}=\frac{1}{4}c_{1\bar{0}}^2+2a_{2\bar{0}}+\frac{1}{16}q_{1\bar{0}}^2\;,\ \ c_{2\bar{0}}=\frac{1}{2}c_{1\bar{0}}^2\;,\nonumber\\&&  d_{2\bar{1}}=2d_0c_{2\bar{1}}+\frac{4}{3}c_{1\bar{0}}d_{1\bar{1}}\;,\ \ \ d_{2\bar{2}}=-5d_{2\bar{0}}+6d_0c_{1\bar{0}}^2-\frac{9}{2}d_0a_{2\bar{0}}\;,\ \ \ b_{3\bar{1}}=-3a_{3\bar{1}}-c_{1\bar{0}}c_{2\bar{1}}\;,\nonumber\\&&
b_{3\bar{3}}=4a_{3\bar{1}}+c_{1\bar{0}}c_{2\bar{1}}\;,\ \ \  c_{3\bar{1}}=c_{1\bar{0}}c_{2\bar{1}}\;,\ \ \ c_{3\bar{2}}=\frac{1}{2}c_{1\bar{0}}^3-\frac{1}{2}c_{1\bar{0}}a_{2\bar{0}}-3c_{3\bar{0}}\;,\ \ \ q_{3\bar{0}}=\frac{1}{6}a_{2\bar{0}}q_{1\bar{0}}\;,\nonumber\\&& d_{3\bar{0}}=-\frac{1}{4}c_{1\bar{0}}d_0a_{2\bar{0}}+\frac{7}{4}d_0c_{3\bar{0}}+\frac{5}{4}c_{1\bar{0}}d_{2\bar{0}}-\frac{9}{8}c_{1\bar{0}}^3d_0-\frac{1}{12}
c_{2\bar{1}}d_{1\bar{1}}\;,\nonumber\\&& d_{3\bar{2}}=\frac{69}{8}c_{1\bar{0}}^3d_0-\frac{27}{4}c_{1\bar{0}}d_0a_{2\bar{0}}-\frac{25}{4}c_{1\bar{0}}d_{2\bar{0}}-\frac{27}{4}d_0c_{3\bar{0}}+\frac{7}{4}c_{2\bar{1}}
d_{1\bar{1}}\;,\nonumber\\&& d_{3\bar{3}}= \frac{22}{3}c_{1\bar{0}}d_0c_{2\bar{1}}-\frac{7}{3}d_{3\bar{1}}-\frac{9}{4}d_0a_{3\bar{1}}-2a_{2\bar{0}}d_{1\bar{1}}+\frac{20}{9}c_{1\bar{0}}^2d_{1\bar{1}}\;,\ \ \ \ \cdot\cdot\cdot\cdot\cdot\cdot\;.
\end{eqnarray}

\newcommand\arctanh[3]{~arctanh.{\bf ~#1}, #2~ (#3)}
\newcommand\ARNPS[3]{~Ann. Rev. Nucl. Part. Sci.{\bf ~#1}, #2~ (#3)}
\newcommand\AL[3]{~Astron. Lett.{\bf ~#1}, #2~ (#3)}
\newcommand\AP[3]{~Astropart. Phys.{\bf ~#1}, #2~ (#3)}
\newcommand\AJ[3]{~Astron. J.{\bf ~#1}, #2~(#3)}
\newcommand\GC[3]{~Grav. Cosmol.{\bf ~#1}, #2~(#3)}
\newcommand\APJ[3]{~Astrophys. J.{\bf ~#1}, #2~ (#3)}
\newcommand\APJL[3]{~Astrophys. J. Lett. {\bf ~#1}, L#2~(#3)}
\newcommand\APJS[3]{~Astrophys. J. Suppl. Ser.{\bf ~#1}, #2~(#3)}
\newcommand\JHEP[3]{~JHEP.{\bf ~#1}, #2~(#3)}
\newcommand\JMP[3]{~J. Math. Phys. {\bf ~#1}, #2~(#3)}
\newcommand\JCAP[3]{~JCAP {\bf ~#1}, #2~ (#3)}
\newcommand\LRR[3]{~Living Rev. Relativity. {\bf ~#1}, #2~ (#3)}
\newcommand\MNRAS[3]{~Mon. Not. R. Astron. Soc.{\bf ~#1}, #2~(#3)}
\newcommand\MNRASL[3]{~Mon. Not. R. Astron. Soc.{\bf ~#1}, L#2~(#3)}
\newcommand\NPB[3]{~Nucl. Phys. B{\bf ~#1}, #2~(#3)}
\newcommand\CMP[3]{~Comm. Math. Phys.{\bf ~#1}, #2~(#3)}
\newcommand\CQG[3]{~Class. Quantum Grav.{\bf ~#1}, #2~(#3)}
\newcommand\PLB[3]{~Phys. Lett. B{\bf ~#1}, #2~(#3)}
\newcommand\PRL[3]{~Phys. Rev. Lett.{\bf ~#1}, #2~(#3)}
\newcommand\PR[3]{~Phys. Rep.{\bf ~#1}, #2~(#3)}
\newcommand\PRd[3]{~Phys. Rev.{\bf ~#1}, #2~(#3)}
\newcommand\PRD[3]{~Phys. Rev. D{\bf ~#1}, #2~(#3)}
\newcommand\RMP[3]{~Rev. Mod. Phys.{\bf ~#1}, #2~(#3)}
\newcommand\SJNP[3]{~Sov. J. Nucl. Phys.{\bf ~#1}, #2~(#3)}
\newcommand\ZPC[3]{~Z. Phys. C{\bf ~#1}, #2~(#3)}
\newcommand\IJGMP[3]{~Int. J. Geom. Meth. Mod. Phys.{\bf ~#1}, #2~(#3)}
\newcommand\IJMPD[3]{~Int. J. Mod. Phys. D{\bf ~#1}, #2~(#3)}
\newcommand\IJMPA[3]{~Int. J. Mod. Phys. A{\bf ~#1}, #2~(#3)}
\newcommand\GRG[3]{~Gen. Rel. Grav.{\bf ~#1}, #2~(#3)}
\newcommand\EPJC[3]{~Eur. Phys. J. C{\bf ~#1}, #2~(#3)}
\newcommand\PRSLA[3]{~Proc. Roy. Soc. Lond. A {\bf ~#1}, #2~(#3)}
\newcommand\AHEP[3]{~Adv. High Energy Phys.{\bf ~#1}, #2~(#3)}
\newcommand\Pramana[3]{~Pramana.{\bf ~#1}, #2~(#3)}
\newcommand\PTP[3]{~Prog. Theor. Phys{\bf ~#1}, #2~(#3)}
\newcommand\APPS[3]{~Acta Phys. Polon. Supp.{\bf ~#1}, #2~(#3)}
\newcommand\ANP[3]{~Annals Phys.{\bf ~#1}, #2~(#3)}
\newcommand\RPP[3]{~Rept. Prog. Phys. {\bf ~#1}, #2~(#3)}
\newcommand\ZP[3]{~Z. Phys. {\bf ~#1}, #2~(#3)}
\newcommand\NCBS[3]{~Nuovo Cimento B Serie {\bf ~#1}, #2~(#3)}
\newcommand\AAP[3]{~Astron. Astrophys.{\bf ~#1}, #2~(#3)}
\newcommand\MPLA[3]{~Mod. Phys. Lett. A.{\bf ~#1}, #2~(#3)}

\end{document}